\documentclass[a4,11pt]{article}
\usepackage{epsfig}
\usepackage{a4,color,graphics,palatino}
\usepackage{graphicx}
\usepackage{amsmath}
\usepackage{bm}% bold math

\def\be{\begin{equation}}\def\ba{\begin{eqnarray}}
\def\ee{\end{equation}}\def\ea{\end{eqnarray}}
\def\ben{\begin{enumerate}}\def\bitem{\begin{itemize}}
\def\een{\end{enumerate}}\def\eitem{\end{itemize}}
\def\no{\nonumber\\}

\def\Tr{{\mbox{Tr}}}
\newcommand{\e}{{\mbox{e}}}

\def\roughly#1{\mathrel{\raise.3ex\hbox{$#1$\kern-.75em%
\lower1ex\hbox{$\sim$}}}}\def\lsim{\roughly<}
\def\gsim{\roughly>}

\def\deriv{\partial}

\begin{document}
\begin{titlepage}
\begin{center}

 \vskip 1.5cm

{\Large \bf  Holographic QCD in medium: a bottom up approach}
\vskip 1. cm
{ Kwanghyun Jo$^{a}$\footnote{e-mail : jokh38@gmail.com},
  Bum-Hoon Lee$^{bc}$\footnote{e-mail : bhl@sogang.ac.kr},
  Chanyong Park$^{b}$\footnote{e-mail : cyong21@sogang.ac.kr},
  Sang-Jin Sin$^{a}$\footnote{e-mail : sjsin@hanyang.ac.kr}}

\vskip 0.5cm

(a) {\it  Department of Physics, Hanyang University, Seoul 133-791, Korea}\\
(b) {\it  Center for Quantum Spacetime (CQUeST), Sogang University, Seoul 127-742, Korea}\\
(c) {\it  Department of Physics, Sogang University, Seoul 121-742, Korea}
\end{center}

\centerline{(\today) }
\vskip 1cm
\vspace{1.0cm plus 0.5cm minus 0.5cm}

\begin{abstract}
A holographic dual of hadrons at finite density is considered. We use the zero black-hole mass limit of Reisner-Nordstrom (RN) AdS background with hard wall to describe a confining background with finite quark density.  We calculate density-dependence of meson masses and decay constants. In our model, pion decay constant and its velocity go down but all the meson masses go up as density grows.
\end{abstract}

\end{titlepage}

\newpage
\setcounter{footnote}{0}
\section{Introduction}
Recently, motivated by the AdS/CFT correspondence \cite{Maldacena}, it has been shown that hadrons can be studied by constructing a 5 dimensional theory where  chiral symmetry is imposed and  the gluon dynamics is encoded  in gravitational warping.
This approach is generically called holographic QCD \cite{SS,EKSS,PR}.
Five dimensional classical theory  gives results of four dimensional hadron physics, fitting well both meson and baryon spectra. The scale in this model is introduced as cut off radius  of the IR region. For the finite temperature case, one has to solve the wave functions in the black hole background and need to impose the infalling boundary conditions for the classical field at the horizon. The result shows that mesons are dissociated resulting in  deconfinement at some high temperature \cite{Holospectral}. Confinement is described by the Hawking-Page transition where thermal AdS is the winning candidate in low temperature over the black hole background \cite{Herzog}.

In this paper we are interested in nuclear matter where the baryonic density is high. One of the characteristic properties of nuclear matter is that
it comes with others. No quarks are coming alone and nucleons in a single nucleus are coming
in multitudes that are in interactions. While properties of hadrons, both mesons and baryons, in medium are fairly accurately understood up to the nuclear matter density thanks to both large amount of experiments performed since many decades and highly sophisticated many-body techniques, there is very little understanding of what happens when nuclear matter is squeezed beyond the density of normal nuclear matter. This is principally because due to basic difficulty in dealing with fermions in the nonperturbative regime of QCD, there are no model-independent inputs available from theory and due to absence -- up to date -- of facilities probing high density, there is little guidance from experiments.
% As will be detailed later, the presently available experimental data are much too ambiguous to be of use for understanding the physics of dense matter. This is why an alternative approach to the issue based on holographic duality which has proven to provide hitherto unforseen information on hot hadronic matter, is appealing.

%Therefore the formalism to treat hadrons in finite dense media is very important.
The holographic approach to encoding the (baryon) chemical potential was  suggested in \cite{FIRST} and has  been
discussed by many authors\cite{density}. While the prescription to the
geometry for the finite temperature has been very clear from the early days of AdS/CFT, it was far from the case for the corresponding geometry for the baryon charge. This is partly due to the fact that the very concept of the fundamental representation was introduced only
by means of ``probe branes."

In Ref.\cite{SSJBFB}, it was suggested that if one introduces the bulk-filling flavor branes,
the gravity back reaction of the baryonic  matter can be easily encoded,
allowing the charge of the RN AdS black hole to be identified with the fermionic charge
as briefly explained in the next section. This back reaction should correspond to the
attractive interaction between the gluon and fermion.
The hydrodynamics and transport coefficients of this system have been analyzed in \cite{SIN2}.
However, what the low-temperature pair for the RN black hole which is the analogue of the
thermal AdS as the pair of AdS black hole could be, has not been clear until very recently.
In \cite{CY}, it was suggested that zero energy gravity solution is the resolution of the puzzle.
There is a naked singularity, but it is hidden inside the hard wall.
This background has the potential that can describe -- for the first time -- density
dependence of a variety of observables in the confining phase.
Especially, the phase diagram for the deconfinement phase transition and the $\rho$-meson
mass were evaluated.
At the present paper, we will make a quantitative study on
other physical observables
in finite baryon density, as well as correcting the numerical error for
the $\rho$-meson mass in the section 4 of \cite{CY}.
Here, the masses and the decay constants of the vector mesons $\rho$, $\rho'$ and the axial
mesons $\pi, a_1, a_2$ are calculated.
We find that the meson masses typically increase at increasing density.
Pion decay constant and its velocity go down but all the meson mass goes up
and as density grows.

\section{The Model}

To describe baryon density in a holographic setup, we will consider the RN AdS black hole geometry as background.
To interpret the charge of the black hole as quark/baryon charge rather than the R-charge, we make use of a simple holographic
model discussed by one of us~\cite{SSJBFB} in which fermions are included as an local U(1) charge  source on the bulk filling branes and  the fermion back reaction on the  metric is included by solving
minimally coupled U(1) source to the gravity.   The holographic model consists of $N_c$ D3 branes with $N_f$
D7 branes touching the D3 branes and filling all of the AdS$_5$ space.   The filling is optimal for D9 branes, but the differences between D7 and D9 are minimal in practical sense.  The induced  metric on D7 with zero current quark mass  is identical to that of AdS$_5$ and the corresponding boundary gauge theory is SYM$_{1+3}$ with $N_f$ massless flavors. We assume that the bulk filling brane tension  does not modify the background metric due to its homogeneity while the fermion charge, which is dual to the local $U(1)$ charge concentrated inside the black hole, requires the back-reaction of the metric.
If we truncate the compact $S^5$ part and consider the gravity coupled with a local charge, the result is the well-known Reissner-Nordstrom AdS (RN-AdS) black hole metric   coupled to a U(1) flavor bulk field  sourced by the fermion charge on the black hole. The point of this construction is that it allows to interpret the local U(1) charge as the U(1) brane charge dual to the global fermion number rather than the R-charge.

To describe the confining phase of this model we introduce a hard wall and set the mass of the black hole to zero,  which corresponds to the zero temperature with finite density.  The metric and
electric potential of charged AdS black hole   background are
\ba
ds^2 &=& \frac{l^2}{z^2} \left(-f(z) dt^2+d\vec{x}^2+\frac{dz^2}{f(z)}\right)\label{RN} \\
f(z) &=& 1+q^2 z^6, \quad A_t=\mu -Q z^2 \nonumber
\ea
where $l$ is the AdS radius. The black hole charge parameter $q$ and the gauge charge $Q$ will be related by the equation of motion it will be related to the quark number density.
We put the fields in the region $0\leq z \leq z_m$ and a hard wall at $z=z_m$. We can interpret this hard wall as a probe D7 brane in the background $N_c$ D3 brane \cite{SSJBFB}. This metric has no horizon  hence has a naked singularity, which however is hidden inside the hard wall, which is typical in most of the confining background.
 We interpret this background as the
low temperature pair of the RN black hole analogous to the thermal AdS as pair of the AdS black hole.
Therefore we propose that it describes the confining phase in finite baryon density following
\cite{CY,Park:2009nb}.

We start with the action  \cite{EKSS}
\be
S = \int d^4x dz \sqrt{-g} \Tr \left[-|D_M X|^2 -M_X^2 |X|^2 - \frac{1}{4 g_5^2} (L_{MN}L^{MN}+R_{MN}R^{MN})\right].
\ee
We assume  the metric is fixed  and given by (\ref{RN}). $X$ is the dual to the quark bilinear operator $ <\bar qq>$ and  $M_X^2 = -3/l^2$  according to the AdS/CFT dirctionary $M_X^2l^2 =\Delta_X(\Delta_X-4)$. And the Tr  is trace over the flavor group. The covariant derivative is
\be
D_M X = \deriv_M X +i L_M X -i X R_M .
\ee
If we write $  X = S \e^{i \pi^a \tau^a}$  with real scalar $S$ and adjoint $\pi^a$, the chiral symmetry breaking is obtained with nonzero vev of $S$  and  $\pi^a$ becomes the Nambu-Goldstone boson of chiral symmetry breaking. Here, $M,N$  run  over 0,1,2,3, z and $\mu,\nu$   run over  0,1,2,3. From L(eft) and R(ight) gauge fields, we form the vector and axial vector gauge fields
\be
V_M = \frac{1}{2}(L_M + R_M), \quad A_M = \frac{1}{2}(L_M - R_M).
\ee
 Note that the length dimensions of the fields are
$[X]=L^{-3/2}, \quad [V]=[A]=L^{-1}, \quad [\pi]=[\varphi]=L^0.
$
%\subsection{Equations of Motion}
\def\w{\omega}
Thanks to the rotational invariance of the boundary theory, i.e., SO(2), we can choose the direction of the wave propagating to be $x^3$, so $k^\mu$ = $(\w,0,0,k)$.  In the axial gauge, $V_z=A_z=0,$  the equations of motion of the vector fields are
\ba
0 &=& \left[\deriv_z^2 +\frac{z f'(z)-f(z)}{z f(z)} \deriv_z+\frac{\w^2}{f(z)^2} \right] V_i(z) \no
0 &=& \left[\deriv_z^2 -\frac{1}{z} \deriv_z + \frac{k^2}{f(z)} \right] V_0(z).\label{vectors}
\ea
There is an extra dimension denoted by $z$,  which  gives rise to the towers of vector and axial vector mesons in 4 dimensions.  Next we  expand  the  5 dimensional vector fields in orthonormal complete bases $h_n$'s and $\alpha_n$'s
\be
V_i(x,z) = \sum_n V_i^{(n)}(x) h_n^V(z), \quad
A_i(x,z) = \sum_n A_i^{(n)}(x) h_n^A(z), \quad
A_0(x,z) = \sum_n A_0^{(n)}(x) \alpha_n^A(z).
\ee
We will consider only the time-like mass, so $V_0$ can be ignored. The wavefunction should be normalized so as to give the proper four-dimensional kinetic term of the boundary gauge theory:
\be
l\int^{z_m}_0 dz \frac{(h_n^V(z))^2}{z f(z)} =1,\quad
l\int^{z_m}_0 dz \frac{(h_n^A(z))^2}{z f(z)} =1,\quad
l\int^{z_m}_0 dz \frac{(\alpha_n^A(z))^2}{z} =1.
\ee
In order to solve these second-order differential equations, we need two boundary conditions:    we take  the Dirichlet  condition at $z=0$, and Neunmann condition at $z_m$, i.e.,
\be V(0)=0, \quad {\rm and }\quad V'(z_m)=0.
\ee
 The equations for the axial-vector fields are
\ba
0 & = & \left[\deriv_z^2 +\frac{z f'(z)-f(z)}{z f(z)} \deriv_z+\frac{\w^2}{f(z)^2} -g_5^2 \frac{v(z)^2 l^2}{z^2 f(z)} \right] A_i(z) \no
0 & = & \left[\deriv_z^2 -\frac{1}{z} \deriv_z -g_5^2 \frac{v(z)^2 l^2}{z^2 f(z)}  \right] A_0(z) \no
0 &= & [\deriv_z^2 - \frac{1}{z}\deriv_z ]\varphi - g_5^2 \frac{v(z)^2 l^2}{z^2 f(z)} (\pi+\varphi) \no
0 & = & m_\pi^2 \deriv_z \varphi + g_5^2 \frac{l^2 v(z)^2 f(z)}{z^2} \deriv_z \pi
\ea
where $v(z)$ is the expectation value of the scalar $S$,
$\langle S\rangle\equiv \frac 12 v(z)$, $i$=1,2 and the time-like mass is defined as $\w^2  = m_n^2$ with $k=0$. Note that $A_{x^3} = \deriv_3 \varphi$ is the longitudinal part of the axial field. Let us define $g_5^2 = \tilde{g}_5^2 l$, then all terms which contain $g_5^2 v^2 l^2$ are expressed as
\be
g_5^2 v(z)^2 l^2 = \tilde{g}_5^2 \tilde{v}(z)^2
\ee
where $\tilde{v}(z)= l^{3/2}v(z)$. The decay constant of the axial mesons and the pion are defined by
\be
F_{a_n}^2 = \frac{l}{g_5^2}\left[{h_n^A}''(z_0)\right]^2, \quad (F_\pi^{t,s})^2 = -\frac{1}{g_5^2} \frac{\deriv_z A_{0,i}^{(0)}}{z}\bigg|_{z_0}.
\ee

\section{Density Dependence of Physical Quantities}

\subsection{Fixing the parameters}
To start with we fix the parameters of the model that can be obtained from what is available. The quark number density is $N_c$ times the baryon density, i.e., $n_q = 3 n_B$ for $N_c$ =3 and the nuclear saturation density is 0.16 fm$^{-3}$, so
\ba
n_0^B &=& \frac{3}{4 \pi r_o^3} = 0.16 \mbox{fm}^{-3} \sim 1.28 \times 10^{-3} \mbox{GeV}^3 \no
n_0^q &=& 3 n^B_0 = 3.84 \times 10^{-3} \mbox{GeV}^3.
\ea
%Note the fact $\hbar$ c = 200 MeV fm.
Normal nuclear matter is in the phase where chiral symmetry is spontaneously broken and color is confined. Hence the first-order phase transition from hadronic to quark-gluon phase predicted in the holographic models~\cite{SSJBFB,CY} must have a critical density above the normal nuclear density.
Quark number density $n_q$  is  related to $Q$ and $q$ through
$\kappa$ and $g_5$  \cite{SSJBFB}:
\be
Q^2 = \frac{3}{2} \frac{g_5^2}{l} \frac{l^3}{\kappa^2} q^2,\quad\quad
n_q = \frac{2 l}{g_5^2} Q = \frac{\sqrt{6}l^2}{\kappa g_5} q .
\ee
We adopt the vacuum parameters fixed by \cite{EKSS}:
\be \label{parameters}
\begin{array}{|c|c|c|}
\hline
\mbox{Input} & \mbox{value} & \mbox{Input}\\
\hline
g_5^2/l & 12\pi^2/N_c&\mbox{Matching OPE of pQCD and hQCD} \\
\hline
m_q&2.0 MeV& m_\pi ~~\mbox{and}~~ f_\pi\\
\hline
\sigma& (323 MeV)^3& m_\pi ~~\mbox{and} ~~f_\pi\\
\hline
z_m&1/(323 MeV)&m_\rho  \\
\hline
\end{array}
\ee
remained is the gravitational constant $\kappa^2$.  We determine $\kappa$  by comparing gravity with QCD thermodynamics:
For QCD, the  free energy  is
\be
{\mathcal{F}}_{G} = \frac{\pi^2}{45} N_c^2 T^4,
\ee
while its gravity  counter part in zero quark density   is
\be
{\mathcal{F}}_{ads} = \frac{l^3}{\kappa^3}\frac{\pi^4 T^4}{2}.
\ee
  By comparing these two,
  we get the relationship between $\kappa$ and $N_c$
\be
\frac{l^3}{\kappa^2} = \frac{2 N_c^2}{45 \pi^2}.
\ee
For $N_c$ = 3, $N_f$ =2
\be \label{nqrelation}
Q^2 = \frac{12 N_c}{15} ~q^2, \quad n_q = \frac{N_c}{3 \pi^2} \sqrt{\frac{N_c}{5}}q \sim 0.0784831~ q.
\ee
In this parametrization, one normal nuclear density is related to q $\sim$ 0.0489278.
Since different authors have different normalization conventions, here we give the summarized below.
\be  \label{tableofconvdiff}
\begin{array}{|c|c|c|c|}
\hline
\mbox{Coeff.} & \mbox{PR}\cite{PR},\cite{Dptihqcdwm} & \mbox{BFB}\cite{SSJBFB} & \mbox{Ours}\\
\hline
l^3/ \kappa^2 & 4 N_c^2 /\pi^2 & N_c^2/4 \pi^2   &  2 N_c^2/ 45 \pi^2 \\
\hline
l/ g_5^2     & N_c/12 \pi^2  & N_c N_f/4 \pi^2 &  N_c/12 \pi^2  \\
\hline
l^2 g_5^2/\kappa^2  & 48 N_c & N_c/N_f & 8N_c/15\\
\hline
\mbox{origin}  & \mbox{VV OPE} & \mbox{SYM free energy}  & \mbox{QCD free energy} \\
\hline
\end{array}
\ee
Note that in \cite{PR}, $N_f$=2, the $N_f$ factor is hidden in Trace operation.
\begin{figure}
\begin{center}
    \includegraphics[angle=0, width=0.40 \textwidth]{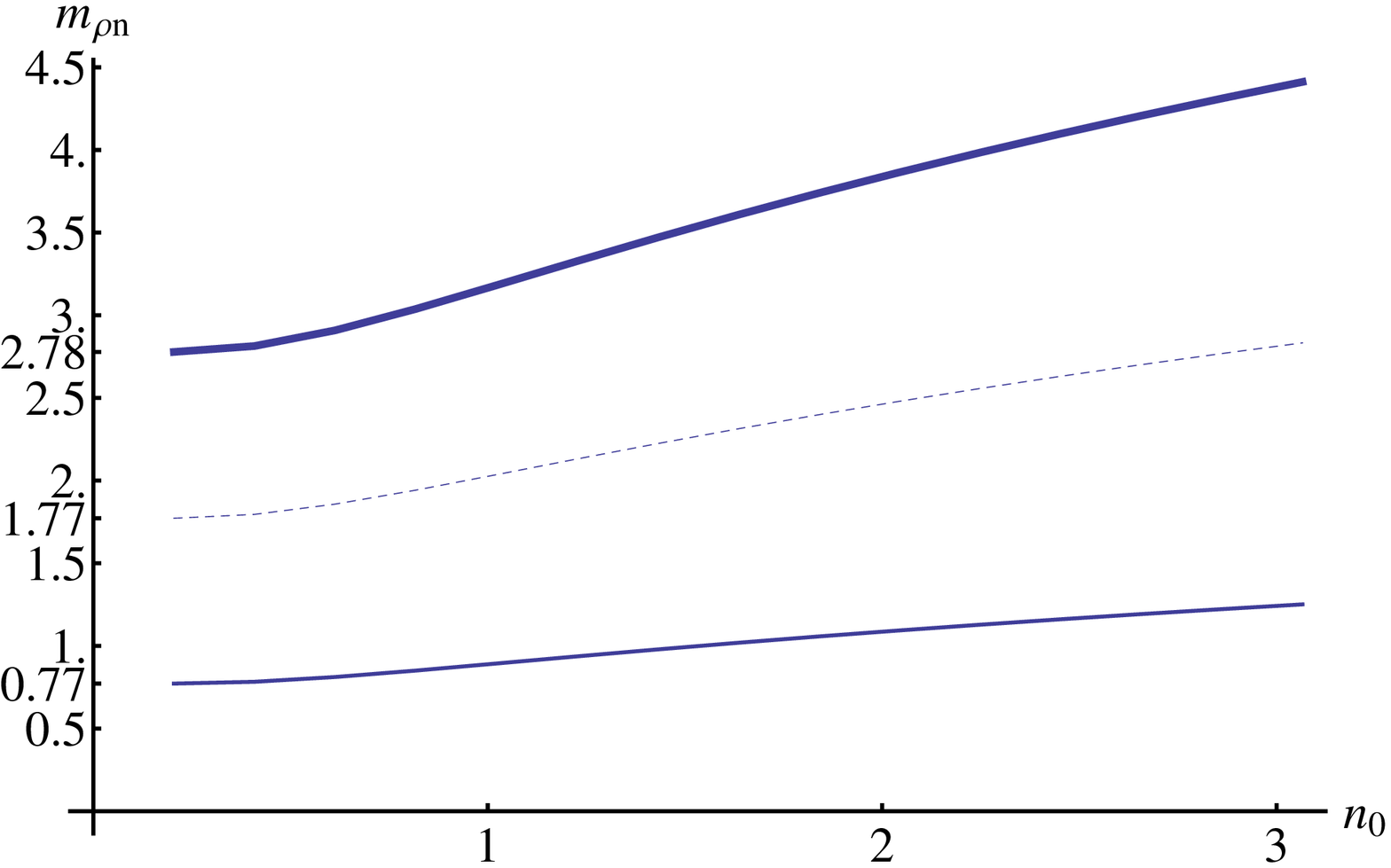}
    \includegraphics[angle=0, width=0.40\textwidth]{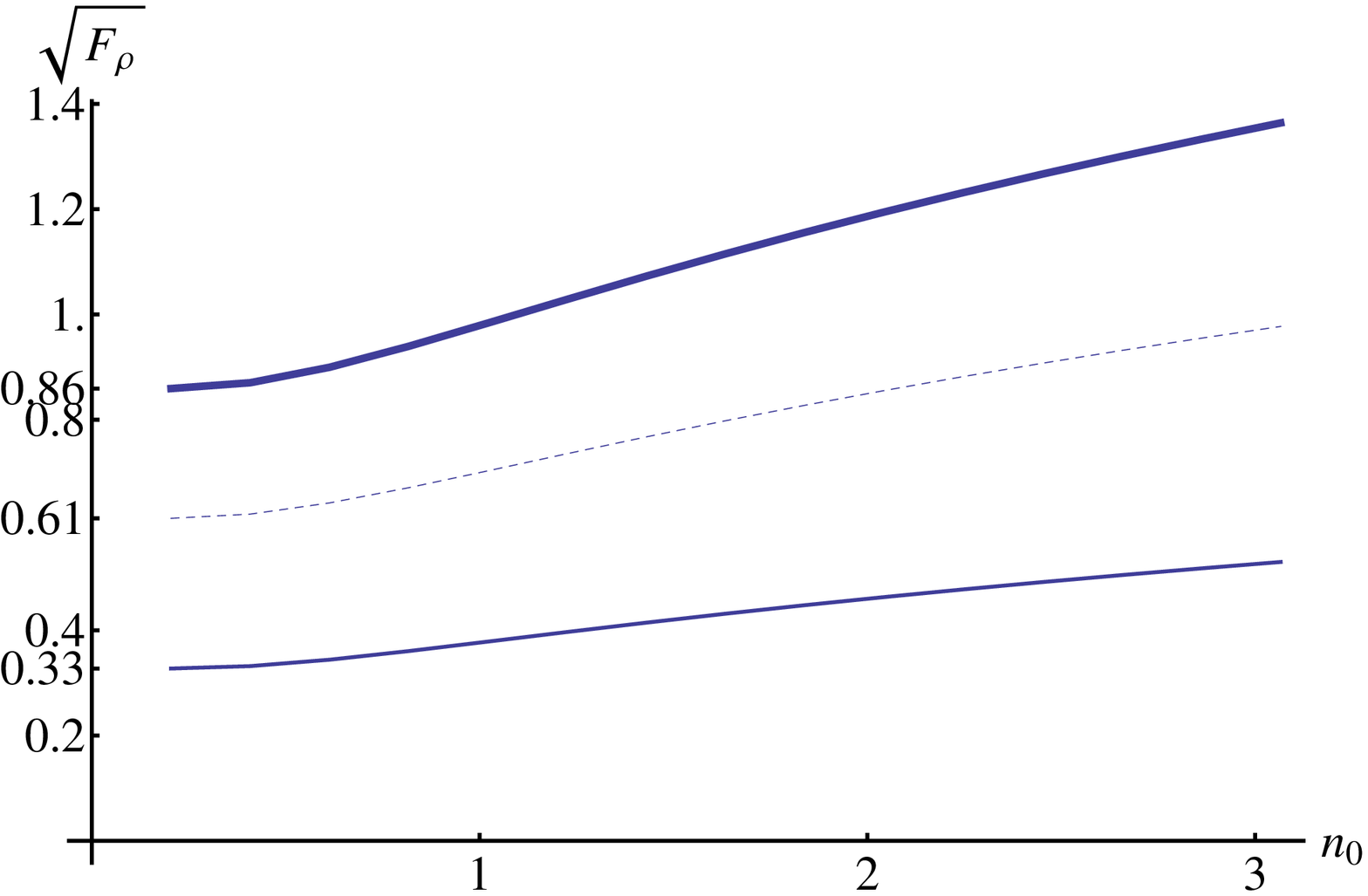}
    \caption{Left:First three $\rho$ meson masses,
    Right:First three rho meson decay constant $\sqrt{F_{\rho_n}}$ in the range 0 $< q <q_c \sim 3~ n_0$}\label{vectormesoncr}
\end{center}
\end{figure}

\subsection{Chiral scalar}
The equation of motion (EOM) for the vacuum expectation value of the scalar field $v(z)$ that comes from $\langle X(z) \rangle = v(z)=l^{3/2} {\tilde v}(z)$   is
\be
\left[\deriv_z^2 +\frac{z f'(z)-3 f(z)}{z f(z)} \deriv_z+\frac{3}{z^2 f(z)} \right] v(z) =0.
\ee
The exact solution for $v(z)$ is
\be
\tilde{v}(z)   = m_q ~z ~{}_2F_1\left(\frac{1}{6},\frac{1}{2},\frac{2}{3},-q^2 z^6\right)+\sigma ~z^3 ~{}_2F_1 \left(\frac{1}{2},\frac{5}{6},\frac{4}{3},-q^2 z^6\right).
\ee
According to the AdS/CFT dictionary, $m_q$ and $\sigma$ correspond to the current quark mass and chiral condensate $\langle\bar{q}q\rangle$, respectively. Following \cite{EKSS}, we take $m_q$=2.29 MeV and $\sigma = (327 MeV)^3$ to reproduce the known values of pion mass and pion decay constant, and $z_m$ = $(323 MeV)^{-1}$ to give the correct $\rho$ meson mass 770 MeV at zero density.
 One of the
disadvantage of hQCD is that there is no way to determine the chiral condensate. It should be provided as an input. In case of top down approach it can be easily calculated by studying the the embedding of the probe brane.

\subsection{Vector and  Axial-vector mesons}
The spectrum of the vector mesons is obtained from the EOM (\ref{vectors}). The procedure is the same as in \cite{EKSS} apart from the $q$ dependence of the mass. For later discussions in confronting nature and comparing with the results of gauge theory models, it is important to understand that this is a ``mean field" approximation in many-body theory language. In fig.~\ref{vectormesoncr}(left), the lowest $\rho$ meson and its first two excitations are shown as a function of density.
All the masses of vector mesons go up.
The decay constants are plotted in figure \ref{vectormesoncr}(right): they also increase at increasing density.

The masses  of axial vector mesons are also increased when density is increased. But unlike vector mesons, increasing mass of axial vector is almost ignorable for lowest mode, $a_1$.  See figure 2.

 \begin{figure}
\begin{center}
    \includegraphics[angle=0, width=0.40\textwidth]{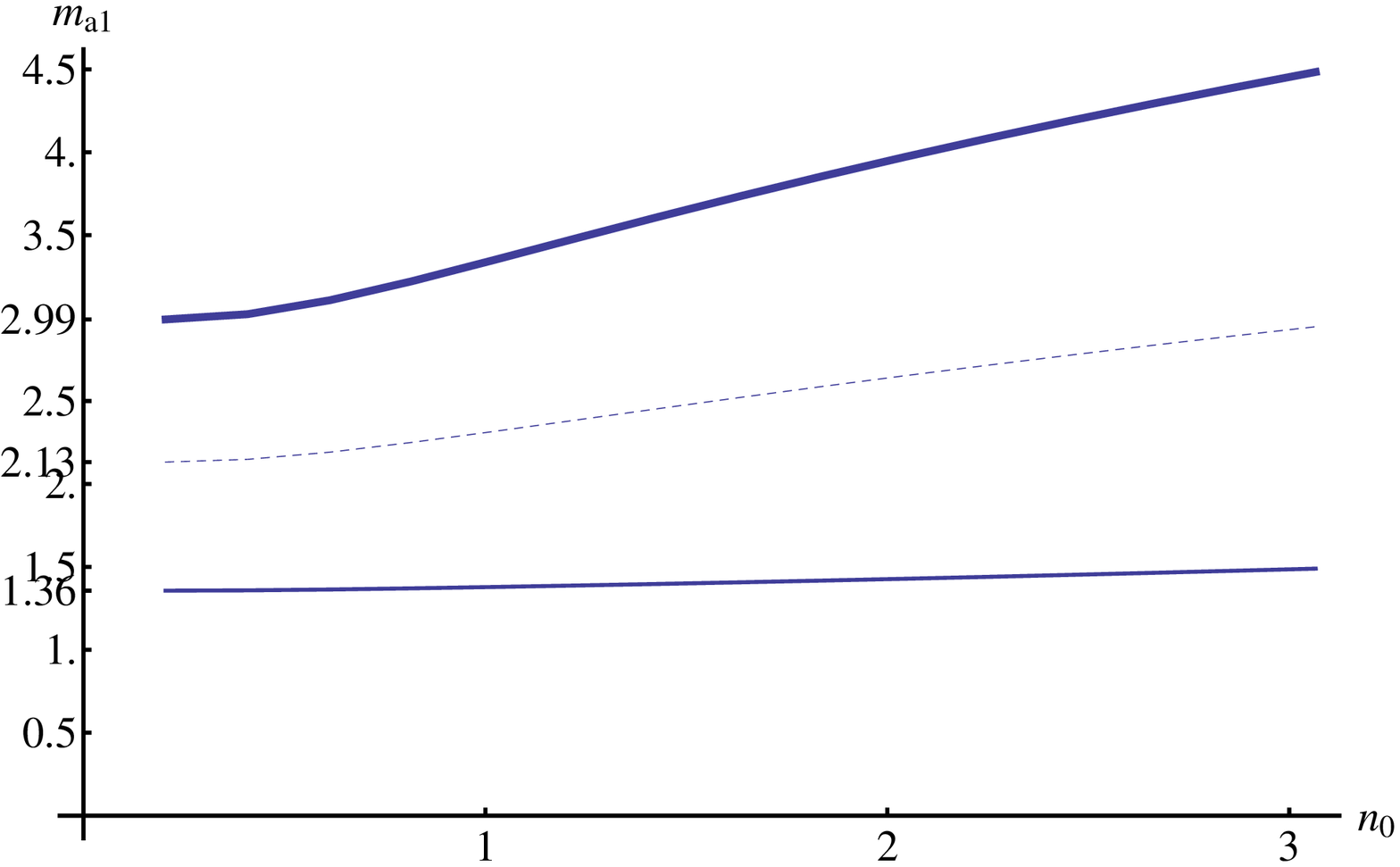}
    \includegraphics[angle=0, width=0.40 \textwidth]{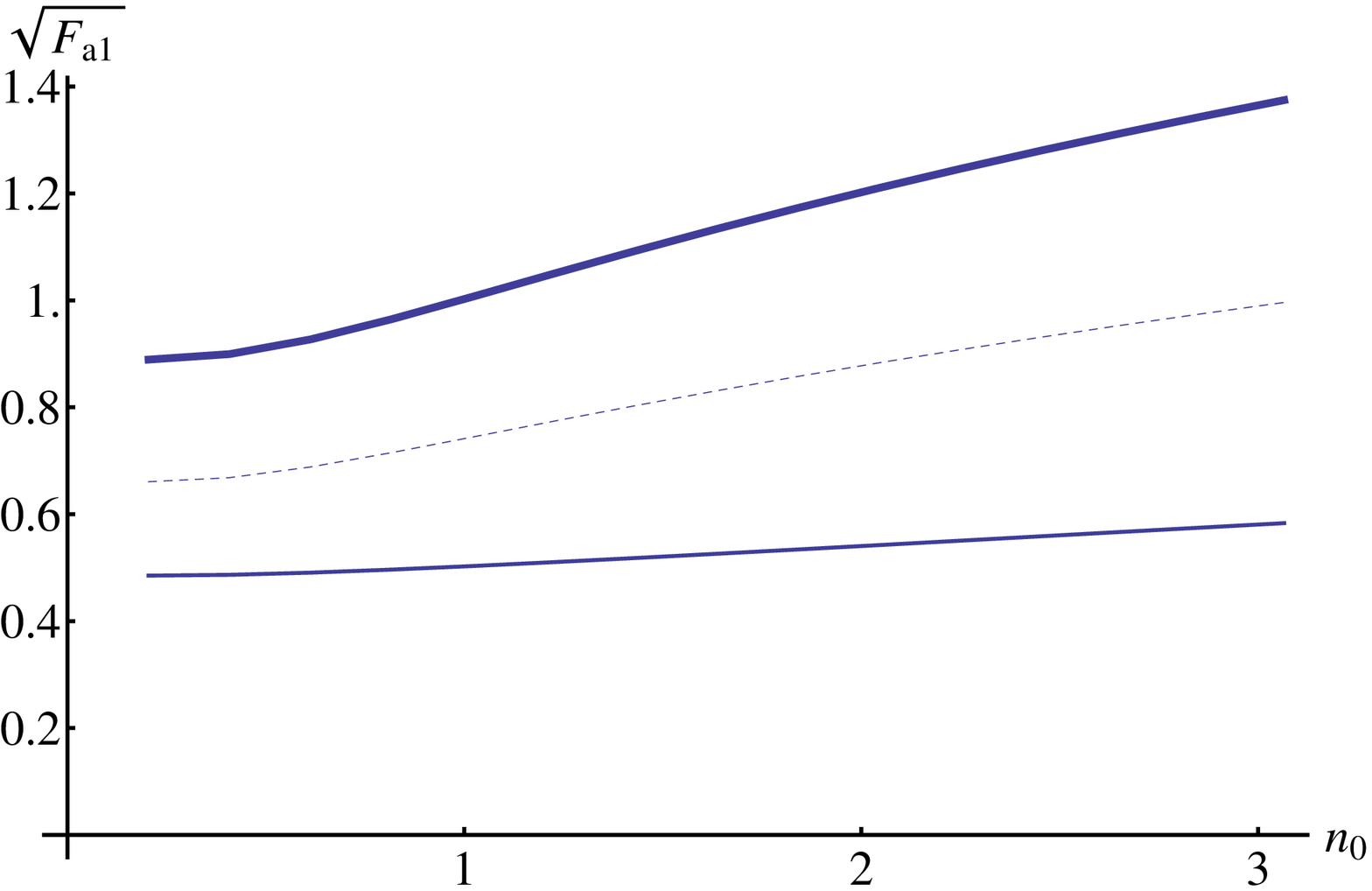}
    \caption{Left: First three axial-vector meson masses,
    Right: First three axial-vector meson decay constants $\sqrt{F_{a_n}}$ in the range 0 $< q <q_c \sim 2.5~ n_0$}\label{axialmeson}
\end{center}
\end{figure}

\subsection{Pion}
Our model shows that the space and time components of the pion decay constant,
$f^i_\pi, f_\pi^t $,  drop at increasing density. At the normal nuclear matter density,
they are reduced to $f_\pi^i \sim 84.3$ MeV and $f_\pi^t \sim$ 91.2 MeV. The pion velocity
defined as the ratio $f^i_\pi/f_\pi^t $ is shown in fig~\ref{sigconpionmassanddecay}.
Both $f^i_\pi$ and $f_\pi^t $ decrease as density increases, as expected from the chiral
symmetry point of view. But the pion mass is increasing.
One does not really know what happens to the pion mass in QCD with proper quark-mass terms.
Experiments in deeply bound pionic atom indicate that the pion mass in medium actually goes
up a few MeV at nuclear matter density. But since  this increase persists in the chiral limit,
this may be considered as a drawback of this model.
Fixing this difficulty may require a nontrivial modification of the model as there seems to
be no easy way to incorporate the ``running" chiral condensate $\sigma$. In the present
treatment, $\sigma$ is introduced as an integration constant of the chiral scalar field
$v(z)$ ignorant of density of the system.

\begin{figure}
\begin{center}
    \includegraphics[angle=0, width=0.5 \textwidth]{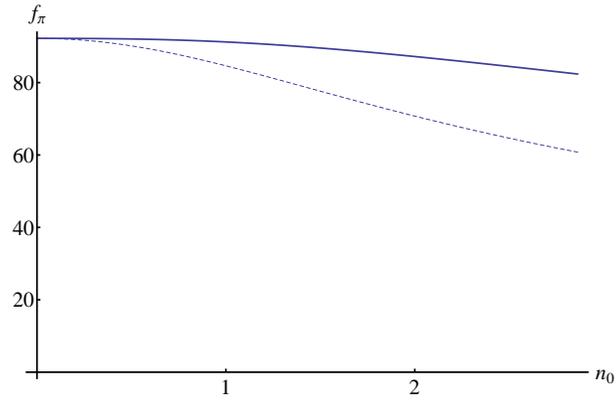}
         \caption{Left: Pion decay constants versus quark number density. The dotted line is for $f_\pi^s$ and the solid line for $f_\pi^t$.}\label{sigconpionmassanddecay}
\end{center}
\end{figure}

\begin{figure}
\begin{center}
     \includegraphics[angle=0, width=0.45 \textwidth]{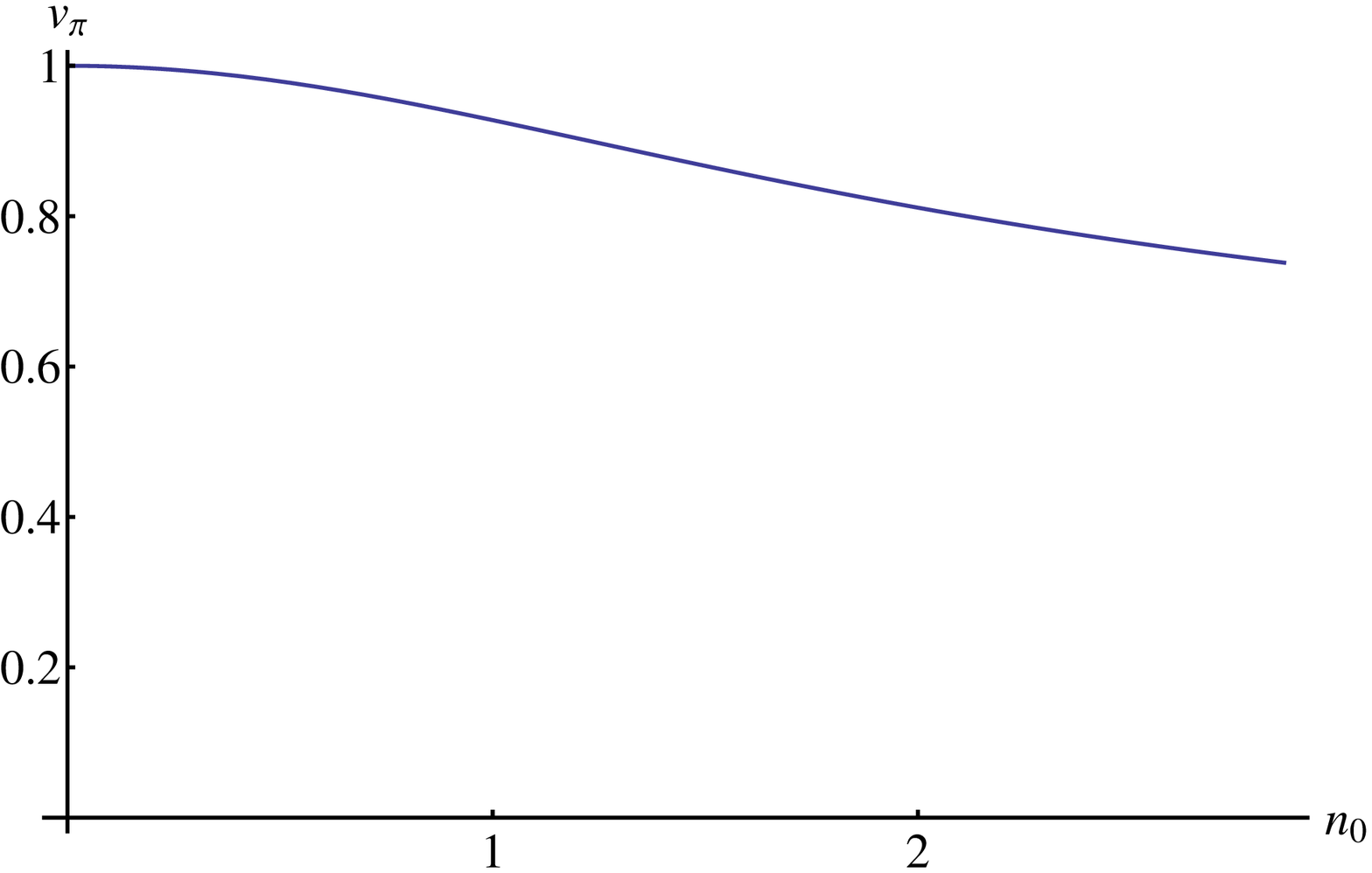}
    \includegraphics[angle=0, width=0.45 \textwidth]{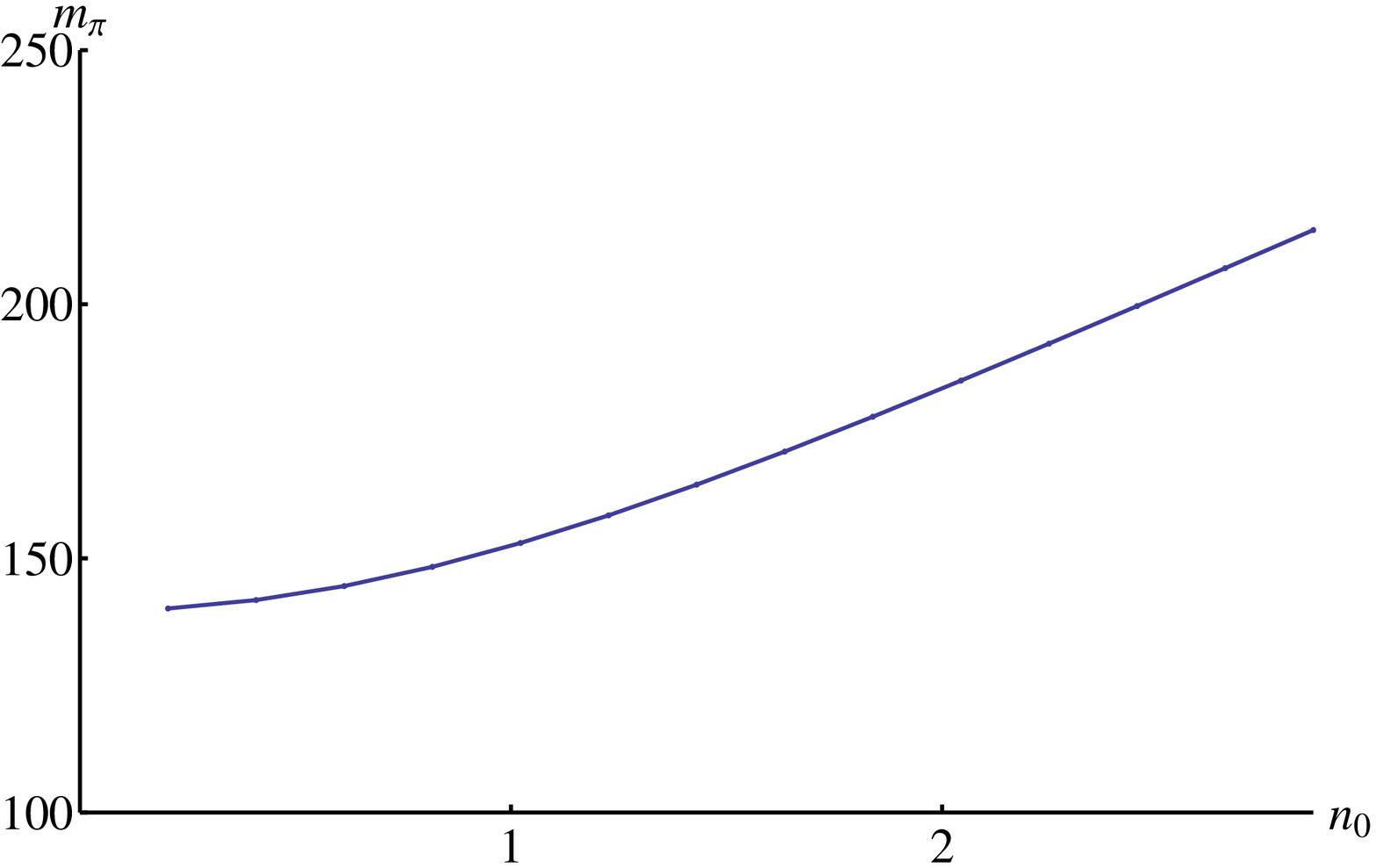}
    \caption{
    Left: Pion velocity, Right: Pion mass}\label{sigconpionmassanddecay}
\end{center}
\end{figure}

\section{Thermodynamics}
It was shown in \cite{CY} that a first-order phase transition between thermal charged AdS and RN AdS black hole takes place. The phase diagram without the calibrating the scale was drawn in the ($\mu$, $T$), ($Q$, $T$) plane, but for different sets of parameters $g_5^2, \kappa^2$. In this section, we will briefly  recast the work of \cite{CY} and draw the phase diagram with determined parameters
in this work, which enables us to compare with real QCD phase diagram.

\subsection{Fixed chemical potential}
The difference of regularized action is given for the fixed chemical potential
\ba
S_{RN} &=& \frac{V_3 l^3}{\kappa^2}\frac{1}{T_{RN}} \left(\frac{1}{\epsilon^4}-\frac{1}{z_+^4}-\frac{2 \kappa^2}{3g_5^2 l^2}\frac{\mu^2}{z_+^2} \right) \no
S_{tc} &=& \frac{V_3 l^3}{\kappa^2}\frac{1}{T_{tc}} \left(\frac{1}{\epsilon^4}-\frac{1}{z_m^4}-\frac{3 \kappa^2}{2g_5^2 l^2}\frac{\mu^2}{z_m^2} \right) \no
\Delta S &=& S_{RN} - S_{tc} = \frac{V_3 l^3}{\kappa^2}\frac{1}{T_{RN}} \left(\frac{1}{z_m^4}-\frac{1}{2 z_+^4}+\frac{3 \kappa^2}{2g_5^2l^2}\frac{\mu^2}{z_m^2}-\frac{ \kappa^2}{3g_5^2l^2}\frac{\mu^2}{z_+^2} \right)
\ea
the last equality comes from identification the Euclidean time periodicity at UV cutoff ($\epsilon \rightarrow 0$). This action is evaluated on physical state with Dirichlet boundary condition to the gauge field at UV, $A_\tau(\epsilon) = i \mu$. When this action difference is zero, there is first order phase transition which is identified to the de/confinement transition in gauge theory side. Suppose that there is a critical point at $z_+ = z_c$, then $\Delta S$ is zero at $z_+ = z_c$. For the convenience, introduce dimensionless quantities,
\be \label{rescalingzmuc}
\tilde{z}_c = \frac{z_c}{z_m}, \quad \tilde{\mu}_c = \mu_c z_m, \quad \tilde{T}_c = T_c z_m,
\ee
then the action difference is rewritten
\be
\Delta S = \frac{V_3 l^3}{\kappa^2}\frac{1}{T_{RN} z_m^4} \left[1-\frac{1}{2 \tilde{z}_+^4}
+\frac{\kappa^2}{g_5^2l^2} \frac{\tilde{\mu}^2}{6\tilde{z}_+^2} \left(
9 \tilde{z}_+^2-2 \right) \right]
\ee
The temperature in dual hadronic background is identified with of RN AdS,
\be
T_{RN} = \frac{1}{\pi z_+}\left(1 - \frac{1}{2}q^2 z_+^6 \right) \no
\ee
and this q is identified to the chemical potential $\mu$
\be
\mu^2 = Q^2 z_+^4 = \frac{3 g_5^2}{2 \kappa^2} q^2 z_+^4
\ee
The critical chemical potential and temperature in fixed chemical potential is
\be
\tilde{\mu}_c = \sqrt{\frac{g_5^2 l^2}{\kappa^2}\frac{3(1-2\tilde{z}_c^4)}{\tilde{z}_c^2 (9 \tilde{z}_c^2-2)}}, \quad
\tilde{T}_c = \frac{1}{\pi \tilde{z}_c}\left(1 - \frac{\tilde{\mu}_c^2}{2} \frac{2 \kappa^2}{3 g_5^2} \tilde{z}_c^2 \right)
\ee
\begin{figure}
\begin{center}
    \includegraphics[angle=0, width=0.45 \textwidth]{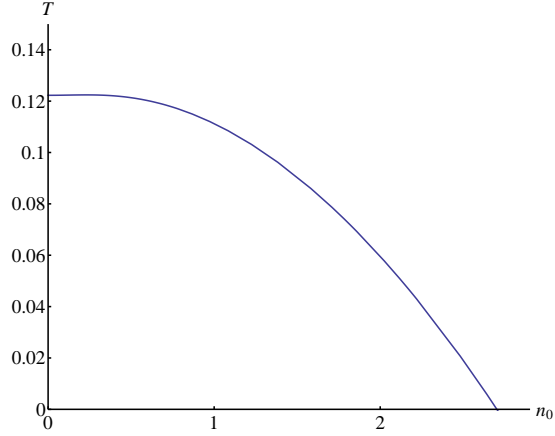}
    \caption{Phase diagram
    %Left : chemical potential vs T , Right
    : number density (in $n_0$ unit) vs T (in GeV unit). }\label{Phasediagram}
\end{center}
\end{figure}

\subsection{Fixed charge}
For the fixed charge case, we should add a boundary term to fix Q. This corresponds to impose Neunmann boundary condition to the at UV, then regularized action of RN AdS is
\ba
\bar{S}_{RN}^D &=& \bar{S}_{RN}^N + S_b = \bar{S}_{RN}^N +\frac{1}{g_5^2}\int_{\deriv \mathcal{M}}d^4 x \sqrt{G^{(4)}}n^MA^NF_{MN} \no
n^M &=& \{0,0,0,0,\frac{z}{l}\sqrt{f(z)} \}, \quad G^{(4)}=\frac{l^4}{z^4}\sqrt{f(z)}
\ea
where $G^{(4)}$ is the determinant of the boundary metric. This $S_b$ is $\mu N$ in statistical sense, so the grand potential is transformed to the Helmholtz free energy by adding this boundary action or imposing Neunmann boundary condition at UV cutoff. The action difference between RN AdS and thermal charged AdS is
\ba
S_{RN}^N &=& \frac{V_3 l^3}{\kappa^2}\frac{1}{T_{RN}} \left(\frac{1}{\epsilon^4}-\frac{1}{z_+^4}+\frac{4 \kappa^2 Q^2}{3g_5^2 l^2} z_+^2 \right) \no
S_{tc}^N &=& \frac{V_3 l^3}{\kappa^2}\frac{1}{T_{tc}} \left(\frac{1}{\epsilon^4}-\frac{1}{z_m^4}+\frac{2 \kappa^2 Q^2}{3 g_5^2 l^2}z_m^2 \right) \no
\Delta S &=& S_{RN} - S_{tc} = \frac{V_3 l^3}{\kappa^2}\frac{1}{T_{RN}} \left[1-\frac{1}{2 \tilde{z}_+^4}+\frac{\kappa^2 Q^2}{3 g_5^2 l^2} \left(5 z_+^2 - 2 z_m^2 \right) \right]
\ea
the critical density and temperature is obtained after rescaling (\ref{rescalingzmuc})
\ba
Q_c z_m^3 & = & \sqrt{\frac{g_5^2 l^2}{\kappa^2} \frac{1-2\tilde{z}_c^4}{\tilde{z}_c^4 (5 \tilde{z}_c^2 - 2)}} \no
\tilde{T}_c &=& \frac{1}{\pi \tilde{z}_c}\left(1 - \frac{\tilde{z}_c^2}{2} \frac{1-2 \tilde{z}_c^4}{5 \tilde{z}_c^2-2} \right)
\ea
Here we set the IR scale $z_m = \frac{1}{0.323} \mbox{GeV}^{-1}$, then $T_c$ = 0.122 GeV.
From the relation eq. (\ref{nqrelation}), the critical density is
\be \label{nQrelation}
n_c=\frac{\sqrt{6}l2}{g_5 \kappa} q_c \sim 1.1 \times 10^{-2} ~ \mbox{GeV}^3 \sim 2.86 n_0
\ee
 The skyrmion approach tells that the critical density is almost two times of normal nuclear density \cite{skyrmionapp}.

\section{Discussion}
Using the zero black-hole mass limit of RN AdS background with its finite charge
identified as the quark (or baryon) density, we studied a holographic dual of dense
baryonic matter. The masses of the vector and axial-vector meson masses and their
decay constants are obtained as a function of density. It is found that all the
meson  mass  go up at increasing density.

We have found that the pion decay constants decrease as density increases and
the pion velocity given by the ratio of the space to time component of the pion decay constant  also decreases.
 Chiral Lagrangian with pions alone  would predict that the pion
velocity goes to zero at $T_c$.
Our model has a glaring defect which needs to be rectified.
The pion mass is found to increase more  rapidly than the inverse of pion decay constant squared, thereby upseting in medium the GMOR relation valid in the vacuum.
Also we can not calculate the density dependence of the chiral condensation from this model, unfortunately.
The cause for these ill behaviors are not known
at the moment. But we conjecture that it is caused by several elements that are
left out in the model.  Even the back reaction of the scalar fields which represent the effects of the gluon and chiral condensates has been ignored. We will report these issues in later works.

\section{Acknowledgement}
We'd like to thanks Mannque Rho for his interest and helpful discussions.
This work was supported in part by the WCU project of Korean Ministry of Education,
Science and Technology (R33-2008-000-10087-0)
 and  by the Korea Science and Engineering Foundation
(KOSEF) grant  through the Center for
Quantum Spacetime(CQUeST) of Sogang University with grant number R11-2005-021. The work of SJS was also supported in part by KOSEF Grant R01-2007-000-10214-0.
The work of KHJ is supported by the Seoul Fellowship.

\newpage

%%%%%%%%%%%%%%%%%%%%%%%%%%%%%%%%%%%%%%%%%%%%%%%%%%%%%%%%%%%
%%%%%%%%%         The References                 %%%%%%%%%%
%%%%%%%%%%%%%%%%%%%%%%%%%%%%%%%%%%%%%%%%%%%%%%%%%%%%%%%%%%%

\end{document}